# Quantum mechanics and the equivalence principle

P.C.W. Davies
Australian Centre for Astrobiology
Macquarie University, New South Wales, Australia 2109

**Abstract**

**A quantum particle moving in a gravitational field may penetrate the classically forbidden region of the gravitational potential. This raises the question of whether the time of flight of a quantum particle in a gravitational field might deviate systematically from that of a classical particle due to tunnelling delay, representing a violation of the weak equivalence principle. I investigate this using a model quantum clock to measure the time of flight of a quantum particle in a uniform gravitational field, and show that a violation of the equivalence principle does not occur when the measurement is made far from the turning point of the classical trajectory. I conclude with some remarks about the strong equivalence principle in quantum mechanics.**

**1. Introduction**

The general theory of relativity, and its plausible near-variants, are founded on the principle of equivalence of inertial and gravitational mass, a property normally associated with Galileo's experiment of dropping different masses from the leaning tower of Pisa. The traditional equivalence principle is fundamentally both classical and local, and it is interesting to enquire how it is to be understood in quantum mechanics. Classically, when the inertial mass $m_i$ and the gravitational mass $m_g$ are equated, the mass drops out of Newton's equations of motion, implying that particles of different mass with the same initial conditions follow the same trajectories. But in Schrödinger's equation the masses do not cancel. For example, in a uniform gravitational field,

$$ - (\hbar^2/2m_i)\partial^2\psi/\partial x^2 + m_g gx\psi = i\hbar\partial\psi/\partial t \qquad (1.1)$$

implying mass-dependant differences in motion. If the motion of a quantum particle is represented by a sharply-peaked wave packet, then by Ehrenfest's theorem one expects that the expectation value of the particle's position will follow a geodesic, and so provide a natural classical limit that complies with the equivalence principle. There will, however, be mass-dependant quantum fluctuations about the mean geodesic motion. This problem has been studied by Viola and Onofrio (1997).

However, one may also consider quantum states of a very different form, for example, energy eigenstates extended over a large region of space. Such quantum states do not have classical counterparts in localised bodies moving on well-defined trajectories. Rather, they might correspond to a steady flux of particles coming from a great distance. What can be said about the principle of equivalence in such a case? In general, a non-local wave function is able to 'feel' the spacetime curvature, and the quantum particle will respond to tidal gravitational forces (Speliotopoulos & Chiao, 2003). But suppose one were to restrict the analysis to a *uniform* gravitational field (as in Eq. (1.1)), for which the spacetime curvature is zero? Might the principle of equivalence then hold even for stationary states?

At first sight the answer would seem to be no. Consider a variant of the simple Galileo experiment, where particles of different mass are projected vertically in a uniform gravitational field with a given initial velocity v. Classically, it is predicted that the particles will return a time $2v/g$ later, having risen to a height $x_{max} = 2(2x/g)^{1/2}$. But quantum particles are able to tunnel into the classically forbidden region above $x_{max}$. Moreover, the tunnelling depth depends on the mass. One might therefore expect a small, but highly significant mass-dependant "quantum delay" in the return time. Such a delay would represent a violation of the equivalence principle.

To investigate this scenario, it is necessary to have a clear definition of the time of flight of the quantum particle. Two problems then present themselves. First, in the case of narrow wave packets one may follow, say, the peak or the median position of the packet as it moves. But this strategy will not work for spread-out energy eigenstates. So how can one measure the time of flight of a particle between fixed points in space when its position uncertainty is very great, without collapsing the wave function to a position eigenstate in the process? Fortunately this problem was solved long ago by Peres (1980), who introduced a simple model quantum clock. The clock measures the *time difference* that a particle takes to travel between two points in space, without disclosing the *absolute* time of passage. This avoids collapsing the wave function to a position eigenstate. The second problem is that, as the particle has a finite probability of penetrating into the classically forbidden region, this analysis involves the vexed issue of how long a particle takes to tunnel into and out of a potential barrier. There is an extensive literature on this contentious topic, which I shall not attempt to summarise here. Rather, I merely remark that the Peres clock yields precise expectation values of tunnelling times that are physically very plausible in the case of square barriers (Davies, 2003), so it seems reasonable to employ this method to discuss the "quantum Galileo experiment."

This topic is of more than purely theoretical interest. It is now possible to observe single atoms moving up and down in the Earth's gravitational field (the so-called atomic fountain experiments), and one could envisage space-based experiments using the much smaller gravitational field of, say, the International Space Station, where the quantum effects are of correspondingly greater relative size.

## 2. Quantum clock

The Peres clock has a single degree of freedom – rotation around a 'clock face' – and is designed to run so long as the moving quantum particle lies within a defined region of space. The expectation value of the 'hand' of the clock remains fixed once the particle has left that region, and may be read by a normal position measurement at any stage subsequently. In effect, this clock measures the change in the phase of the wave function, $\Delta\theta$, with the elapsed time given by

$$\Delta T = \hbar(\partial/\partial E)\Delta\theta. \qquad (2.1)$$

Full details are given in Peres (1980).

As an illustration of the use of the Peres clock for a simple out-and-back time measurement, consider reflection from a square potential step in one space dimension:

$$V = 0, \quad x < 0$$
$$V = V_o, \quad x > 0. \tag{2.2.}$$

The Schrödinger equation for a uniform flux impinging on the step from the left with $E < V_o$ may readily be solved to give

$$u(x) \propto e^{ikx} + Ae^{-ikx}, \quad x<0$$
$$\propto Be^{-px}, \quad x>0, \tag{2.3}$$

where $k = (2mE/\hbar^2)^{1/2}$, $p = [2m(V_o - E)/\hbar^2]^{1/2}$, and $A$ and $B$ are constants. Continuity of $u$ and $u'$ at $x = 0$ yields

$$A = -(p + ik)/(p - ik). \tag{2.4}$$

At a point $x = -d$ from the step, the phase change between the incident (right-moving) and reflected (left-moving) wave is

$$\Delta\theta = 2kd + \arctan(\mathcal{I}m\,A/\mathcal{R}e\,A). \tag{2.5}$$

The latter term on the right of Eq. (2.5) represents the contribution from the particle tunnelling into the classically forbidden region $x > 0$. Using Eq. (2.1) one obtains:

$$\Delta T = 2d/v + 2a/v \tag{2.6}$$

where $v = \hbar k$ corresponds to the classical velocity and $a \equiv 1/p$ is a measure of the depth of penetration of the evanescent wave into the barrier. (Note that it is not necessary to determine the overall normalization factor of the wave function to compute $\Delta T$; any constant phase factor will be eliminated by the differentiation in Eq. (2.1).) The expectation value for the time of flight therefore corresponds to the classical time for the particle to reach the barrier a distance $d$ away at speed $v$, and then return at the same speed, plus an extra duration for it to tunnel beyond the classical reflection point. This 'quantum delay' in the reflection time may be written

$$2\hbar/[E(V_o - E)]^{1/2} = (2/m)^{1/2}(\hbar/v)(V_o - \tfrac{1}{2}mv^2)^{-1/2} \tag{2.7}$$

which vanishes as expected in the classical limit, $\hbar \to 0$, and for an infinite step, $V_o \to \infty$. Note that for a fixed $v$ this expression is $m$-dependant.

Although it is not meaningful to assign a velocity to the particle beneath the barrier, in this simple example the Peres clock yields a tunnelling duration which is in fact equal to the barrier penetration depth divided by the classical speed of the incident particle.

As a second example, consider the potential

$$V = \alpha e^{\beta x}. \tag{2.8}$$

The Schrödinger equation may be solved for the following energy eigenstates

$$u(x) \propto e^{\pi k/\beta} K_{2ik/\beta} \left(\sqrt{(8m\alpha/\hbar^2\beta^2)}e^{\beta x/2}\right) \tag{2.9}$$

which are chosen to remain finite as $x \to \infty$. Evaluating Eq. (2.9) in the limit $x \to -\infty$, where $V \to 0$, yields

$$\pi(\beta/k)e^{\pi k/\beta} \text{cosech}(2\pi k/\beta)\{\Gamma^{-1}(2ik/\beta)\exp[ikx + i(k/\beta)\ln(2m\alpha/\hbar^2\beta^2)] + \text{cc}\}. \tag{2.10}$$

The first term in the outer brackets represents a wave travelling to the right, approaching the potential hill, and the complex conjugate term represents the reflected wave. The phase change between the incident and reflected waves at $x = -X < 0$ is

$$\Delta\theta = -2kX - 2\varphi - 2(k/\beta)\ln(2m\alpha/\hbar^2\beta^2) \tag{2.11}$$

where

$$\varphi = \arctan[\mathcal{I}m(\Gamma^{-1})/\mathcal{R}e(\Gamma^{-1})]. \tag{2.12}$$

The phase factor $\varphi$ simplifies for large $\beta$:

$$\varphi \approx \arctan(\beta/2kC) \tag{2.13}$$

where $C$ is Cantor's number. Using Eqs. (2.1) and (2.11) we obtain for the out-and-back time of flight

$$\Delta T = -2X/v + (2/\beta v)\ln(4E/\alpha) + (4/\beta v)[-\ln(2mv/\hbar\beta) + C/(1 + 4C^2m^2v^2/\hbar^2\beta^2)] \tag{2.14}$$

where $v \equiv \hbar k/m$, this being the incident classical velocity in the asymptotic region $x \to -\infty$, where $V \to 0$. The first two terms on the right of Eq. (2.14) is the classical time of flight; the third term is the quantum correction that takes account of tunnelling. Note that this correction vanishes as $\beta \to \infty$, as expected. In this case the potential becomes a sharp wall, and the situation is identical to the previous example with $V_o \to \infty$. In this limit, the quantum correction is clearly mass-dependant (for fixed $v$) and positive.

**3. The gravitational case**

I now consider the situation for a quantum particle moving in a uniform gravitational potential in one dimension:

$$V(x) = m_g gx. \tag{3.1}$$

Solutions of the Schrödinger Eq. (1.1) that are finite for $x \to \infty$ are well-known in terms of Airy functions (Bessel functions of order 1/3) (see, for example, Davies & Betts, 1994). For eigenstates of energy E one obtains

$$u(x) \propto Ai\,[(x-b)/a] \tag{3.2}$$

where

$$a = (\hbar^2/2m_i m_g g)^{1/3}, \qquad b = E/m_g g, \tag{3.3}$$

and from here on, unless specified explicitly, I shall put $m_i = m_g \equiv m$. The solution is shown in Fig. 1. Note that the particle may tunnel into the classically forbidden region $x > b$, where $Ai$ decays exponentially (evanescent wave) to a depth of order $a$. For an electron near the Earth's surface, this distance is about 1mm. Near an object such as a space station, this tunnelling distance would be several orders of magnitude greater.

We wish to consider the wave function in the region $x < 0$ where the following decomposition in terms of Bessel functions applies:

$$Ai\,(-z) = \tfrac{1}{3}\sqrt{z}\,[J_{1/3}(\zeta) + J_{-1/3}(\zeta)], \tag{3.4}$$

where

$$\zeta = \tfrac{2}{3}z^{3/2}, \qquad z \equiv (b-x)/a > 0.$$

The right hand side of Eq. (3.4) may be re-expressed as the linear combination

$$\tfrac{1}{3}\sqrt{z}\{[e^{i\pi/3}J_{1/3}(\zeta) + e^{-i\pi/3}J_{-1/3}(\zeta)] + [(1 - e^{i\pi/3})J_{1/3}(\zeta) + (1 - e^{-i\pi/3})J_{-1/3}(\zeta)]\}. \tag{3.5}$$

The second term in the outer parentheses is in fact just the complex conjugate of the first. The terms in the square brackets correspond to incident (up-moving) and reflected (down-moving) waves respectively, as may be verified by computing the current vectors of each:

$$j(x) = (\hbar/2im)(u^*du/dx - cc) = \pm (N^2/4\pi)(2\hbar/m)^{1/3} = \text{constant}, \tag{3.6}$$

where $N$ is a normalisation constant, and use has been made of the Wronskian relation

$$J_{1/3}(\zeta)\,J'_{-1/3}(\zeta) - J_{-1/3}(\zeta)\,J'_{1/3}(\zeta) = (2/\zeta\pi)\sin(\pi/3). \tag{3.7}$$

The decomposition into incident and reflected waves may also be seen explicitly by taking the large $z$ limit of the Bessel functions:

$$Ai(-z) \approx (1/3\pi)^{1/2}z^{-1/4}[e^{i\pi/3}\cos(\zeta - 5\pi/12) + e^{-i\pi/3}\cos(\zeta - \pi/12)] + cc. \tag{3.8}$$

After some work, one finds

$$Ai(-z) \approx \tfrac{1}{2}\pi^{-1/2}z^{-1/4}e^{i(\zeta-\pi/4)} + \tfrac{1}{2}\pi^{-1/2}z^{-1/4}e^{-i(\zeta-\pi/4)}$$

$$\approx \pi^{-1/2}z^{-1/4}\sin(\zeta + \pi/4) \tag{3.9}$$

which contains incident and reflected waves of equal amplitude but different phase.

Consider now the phase change between the incident and reflected waves at $x = -X < 0$ far from the classical turning point, i.e. in the above limit of $X \to \infty$, consequent upon the reflection and propagation:

$$\Delta\theta = 2(\zeta + \pi/4) = 4/3[(E/mg - X)/a]^{3/2} - \pi/4. \tag{3.10}$$

From this one obtains, using Eq. (2.1), the out-and-back (or up-and-down) travel time

$$\Delta T = (2\hbar/mga)[(E/mg - X)/a]^{1/2} = 2[2(b - X)/g]^{1/2}. \tag{3.11}$$

A classical particle with total energy $E$ projected vertically at $x = -X < 0$ will climb to a distance $d \equiv b + X$ in a time $\sqrt{(2d/g)}$, so its turnaround time is

$$\Delta T_{\text{classical}} = 2\sqrt{(2d/g)}. \tag{3.12}$$

Comparing Eqs. (3.11) and (3.12) we see immediately that they are the *same*: the expectation value for the turnaround time of a quantum particle is identical to the classical time, when the measurement is performed far from the classical turning point. In this sense, the principle of equivalence holds even for a quantum particle in an unbound delocalised energy eigenstate.

How can this result be understood, when we know that that the wave function is non-vanishing beyond the classical turning point $x = b$? Subject to the usual caveats about attempting to interpret quantum processes in classical language, a plausible explanation is apparent. Whilst there is a finite probability that a given particle may tunnel into the region above the classical turning point and return late, there is also a finite probability that the particle may back-scatter off the gravitational potential before it reaches $x = b$. This is consistent with the fact that the wave function (hence probability density) dips prior to $x = b$ (see Fig. 1). This may be contrasted with the classical probability density, which rises like $(x - b)^{-1/2}$ near $x = b$ for a uniform stream of particles. It would appear that, in the case of the uniform gravitational potential Eq. (3.1), these two effects exactly cancel, leading to neither a shortening nor a lengthening of the classical turnaround time due to quantum effects.

This state of affairs pertains only to large distances from the classical turning point. If one were to measure the turnaround time close to $x = b$, one might expect to "miss" those particles that have scattered back early, and to record a positive delay due to quantum penetration of the gravitational potential. This interpretation is confirmed by computing the phase change just below the classical turning point at $x = b$. Expanding the Bessel functions for small $\zeta$ in Eq. (3.4) yields

$$3^{-2/3}\Gamma^{-1}(2/3)e^{i\pi/3}z \qquad \text{and} \qquad 3^{-4/3}\Gamma^{-1}(4/3)e^{-i\pi/3} \tag{3.13}$$

for the incident and reflected waves respectively, from which the phase change is found to be

$$\Delta\theta = \arctan\{2\sqrt{3}[3^{2/3}\Gamma(2/3)z - 3^{4/3}\Gamma(4/3)]/[3^{2/3}\Gamma(2/3)z + 3^{4/3}\Gamma(4/3)]\}. \quad (3.14)$$

It is of interest to reinstate the distinction between $m_i$ and $m_g$ at this stage. Using Eq. (2.1) and then taking $z \to 0$ afterwards gives

$$\Delta T = 4.3^{-1/6}\Gamma(2/3)\hbar/13\Gamma(4/3)m_g g a \approx 0.5(\hbar m_i/m_g^2 g^2)^{1/3} \quad (3.15)$$

which indeed shows a non-zero tunnelling delay time.

A similar result may be obtained by calculating the so-called dwell time, defined to be the probability $P$ of the particle residing in the classically forbidden region divided by the incident flux (so that the normalisation constants cancel). In the region $x > b$, $z < 0$, the Airy function may be written in terms of a MacDonald function $K_{1/3}$, whence

$$P = (N^2 a/3\pi^2) \int_0^\infty z K_{1/3}^2(\zeta) d\zeta \quad (3.16)$$

where now $\zeta = 2/3(-z)^{3/2}$. The integral may be evaluated:

$$P = (N^2 a) 3^{1/3} \Gamma^2(2/3)/4\pi^2. \quad (3.17)$$

Dividing $P$ by the flux given by Eq. (3.6) yields a value for the dwell time beneath the potential barrier of

$$(1/\pi)(3/4)^{1/3}\Gamma^2(2/3)(\hbar m_i/m_g^2 g^2)^{1/3} \approx 0.4(\hbar m_i/m_g^2 g^2)^{1/3}. \quad (3.18)$$

Comparing Eqs. (3.15) and (3.18) we see that the dwell time is very close to the expectation value of the tunnelling time as determined by the Peres clock. To obtain an idea of the numbers involved, for an electron near the Earth's surface, the dwell time in the tunnelling region is about 4 ms – significantly long by the standards of atomic physics.

## 4. Discussion

The results of this paper suggest that a uniform gravitational potential – which applies locally to any non-singular gravitational field – has a special property in relation to quantum mechanics, namely, that the expectation time for the propagation of a quantum particle in this background is identical to the classical propagation time. This may be taken as an extension of the principle of equivalence into the quantum regime (for a broader discussion of what is entailed by a 'Quantum Equivalence Principle,' see Lämmerzahl, 1996). This special property seems to depend on the form of the potential; it does not apply in the case of a sharp potential step, or an exponential potential, for example.

The result applies only to measurements made far from the classical turning point at $x = b$. The distance scale for this approximation is determined by the length $a = (\hbar^2/2m^2g)^{1/3}$, which roughly corresponds to one de Broglie wavelength from the turning point. Within this distance there are significant quantum corrections to the turnaround time, including the possibility of a mass-dependant delay due to the penetration of the classically forbidden region $x > b$ by the evanescent part of the wave function (3.2). Thus quantum 'smearing' of the equivalence principle is restricted to distances within the normal position uncertainty of a quantum particle. It is noteworthy, however, that even in the Earth's gravitational field the quantum corrections near the classical turning point are considerable. For an electron, the average penetration depth into the gravitational potential is about 1 mm and the corresponding delay time is several milliseconds. It is conceivable that such effects may be measurable using existing technology. In a space-based experiment, utilising the small gravitational field of the space vehicle, the effects would be very much greater.

Because the Peres clock is itself a quantum system, it is subject to intrinsic uncertainty in its performance. There will also be a back-reaction of the operation of the clock on the measured particle (Peres, 1980). These effects introduce errors in $\Delta T$ comparable to Eqs. (3.15) and (3.18). The back-reaction may be reduced by making the coupling between the clock and the particle arbitrarily small, but at the expense of increasing the uncertainty in the clock position. However, the latter uncertainty may be compensated by introducing a large ensemble of identical systems, and regarding the time measurement as a *weak measurement* (Aharanov et. al., 1988). It is in this ensemble sense that the times computed in this paper are to be regarded.

I have restricted attention to the so-called weak equivalence principle. One might also enquire into the status of the strong or Einstein equivalence principles in quantum mechanics. Einstein made the postulate that all of physics in a uniform gravitational field should be locally equivalent to the physics in a uniformly accelerated frame. Does it apply to the problem discussed in this paper? It is well-known (see, for example, Viola & Onofrio, 1997) that under a transformation of coordinates to an accelerated reference frame

$$x' = x - vt - \tfrac{1}{2}at^2$$
$$t' = t$$
(4.1)

the Schrödinger equation for a free particle is transformed to a Schrödinger equation for a particle moving in a uniform gravitational potential with $g = a$. So there is a formal correspondence between a uniform gravitational field and a uniform acceleration in the underlying quantum kinematics, just as there is in classical kinematics. However, in quantum mechanics the relationship between the state of the system and the dynamical evolution is much more subtle than in classical mechanics. In the case of states that represent localised wave packets, the equivalence of acceleration and gravitation goes through in a reasonably straightforward manner (Viola & Onofrio, 1997). But what about the case of the non-localised energy eigenstates considered in this paper? Here the situation is more complicated. The stationary states of the Hamiltonian

$$-(\hbar^2/2m)\partial^2/\partial x^2 + mgx$$
(4.2)

have the form

$$Ai[(x-b)/a]e^{-iEt/\hbar} \qquad (4.3)$$

for energy *E*. Now consider the situation viewed from an accelerated frame, where the particle now moves freely. The appropriate energy eigenstates have the form

$$e^{ikx - iEt/\hbar}. \qquad (4.4)$$

Under the transformation (4.1), the eigenfunctions (4.4) do not transform into (4.3). The equation of motion may transform correctly, but the energy eigenstates do not. Rather, the Airy functions will be complicated linear combinations of plane wave solutions (4.4) and their complex conjugates. (The transformation of plane wave solutions into accelerated reference frames is a well-studied problem; see, for example, Birrell & Davies (1982), section 4.5.) This would not matter if the results of the analysis were linear in the wave function. That is indeed the case for the behaviour of wave packets which are made up of linear combinations of plane waves. But it is not the case for a measurement of the transit time, at least when such a measurement is made using the Peres clock prescription considered here. That is because the time interval depends on a measurement of the phase change, and the sum of the phases of a superposition of waves is generally not the same as the phase of the sum. A Peres clock will generally respond to a superposition of states in a very complicated way. To be sure, a stationary Peres clock will register the classical transit time $\Delta T = d/v$, where $v = \hbar k$ and $d$ is the traversed distance, for the energy eigenstate (4.4). But one cannot use this simple fact combined with the transformation (4.1) to deduce the main result Eq. (3.11). To test the quantum correspondence between stationary and accelerated reference frames, it is necessary to consider the response of an *accelerated* quantum clock to the state (4.4). Given the well-known differences between the results of alternative model clocks to quantum tunnelling times, and given the small but significant difference found here between Eqs. (3.15) and (3.18), it is far from clear that all model quantum clocks will respond equally in this scenario. I shall report on this topic in a future publication.

**Acknowledgements**

I should like to thank Raymond Chiao and Dipankar Home for helpful discussions.

**References**

Aharaonov, Y., D.Z. Albert & L. Vaidman (1988) *Phys. Rev. Lett*. **60**, 1351.
Chiao, R
Davies, P.C.W. & David Betts (1994), *Quantum Mechanics*, second edition (Chapman & Hall), p. 26.
Davies, P.C.W. & N.D. Birrell (1982) *Quantum Fields in Curved Space* (Cambridge University Press).

Davies, P.C.W. (2003) 'Can a quantum particle tunnel faster than light?' unpublished preprint.
Lämmerzahl, C. (1996) *Gen. Rel. Grav*. **28**, 1043.
Peres, A. (1980) *Am. J. Phys*. **48**, 552.
Speliotopoulos, A.D. & R.Y. Chiao, 'Coupling to linearized gravity: dynamics in the general laboratory frame," *Phys. Rev. D*, in the press.
Viola, L. & R. Onofrio (1997) *Phys. Rev. D* **55**, 455.

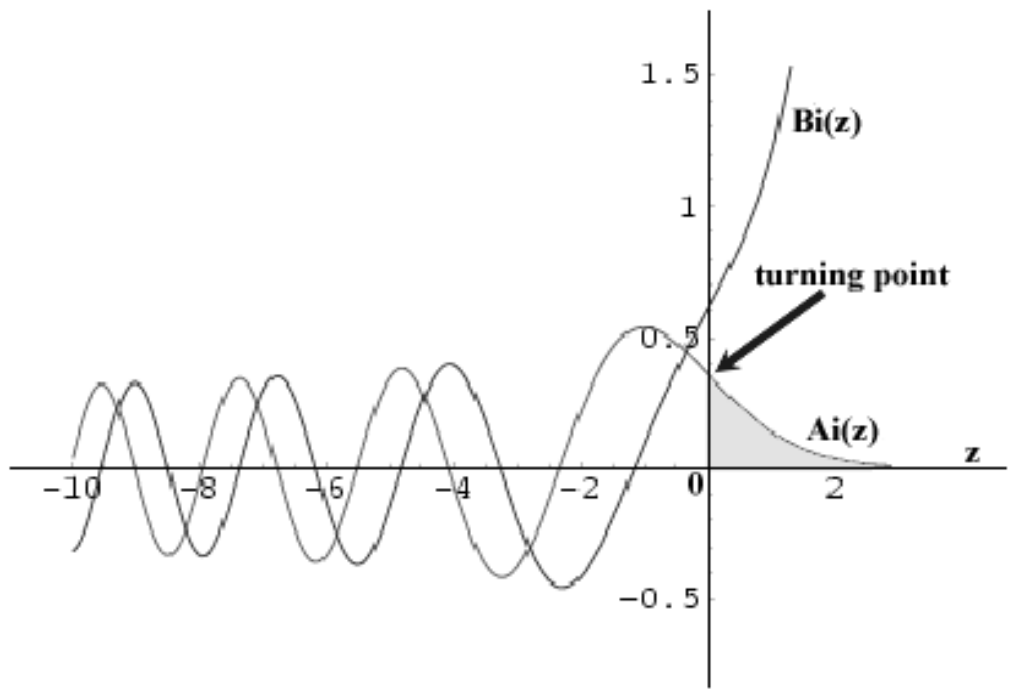

Fig. 1

**Caption**

The real function Ai(z) is bounded as z → ∞, and so is chosen as the appropriate wave function in place of the linearly independent solution Bi(z), which is unbounded as z → ∞. The point z = 0 corresponds to the classical turning point of the particle's trajectory. Note that Ai(z) oscillates in the region z < 0, and dips just before z = 0, indicating that there is a finite probability of the particle scattering back before reaching the classical turning point. Conversely, there is a non-zero probability that the particle will be found in the classically forbidden (shaded) region z > 0.